\newif\ifjournal  
\newif\ifdraft    

\drafttrue

\ifjournal
  \ifdraft
    \documentclass[acmsmall, review=true]{acmart}
  \else
    \documentclass[acmsmall]{acmart}
  \fi
\else
  \documentclass[times]{article}

  \usepackage{geometry}
  \usepackage{natbib}
  \usepackage{url}
\fi

\usepackage{algorithm2e}
\usepackage{amsmath}
\usepackage{array}
\usepackage{booktabs}
\usepackage{caption}
\usepackage[noabbrev]{cleveref}
\usepackage{color}
\usepackage{enumerate}
\usepackage{float}
\usepackage{fontawesome}
\usepackage{graphicx}
\usepackage{mathbbol}
\usepackage{moreverb}
\usepackage{multirow}
\usepackage{standalone}
\usepackage{stmaryrd}
\usepackage{subcaption}
\usepackage{tikz}
\usepackage{xcolor}
\usepackage{listings}

\usepackage[colorinlistoftodos,prependcaption,textsize=small]{todonotes}
\usepackage{xargs}
\usepackage{soul}
\definecolor{RedOrange}{cmyk}{0, 0.77,0.87,0}
\newcommandx{\fixap}[1]{\todo[linecolor=green,backgroundcolor=green!25,bordercolor=green,inline]{\textbf{AP: }#1}}
\newcommandx{\fixdlg}[1]{\todo[linecolor=green,backgroundcolor=green!25,bordercolor=green,inline]{\textbf{DLG: }#1}}
\newcommand{\code}[1]{\texttt{#1}}
\newcommand{\UP}[1]{#1}

\makeatletter
\def\blfootnote{\xdef\@thefnmark{}\@footnotetext}
\makeatother

\definecolor{Tred}   {RGB}{255, 152, 150}
\definecolor{Tgreen} {RGB}{152, 223, 138}
\definecolor{Tblue}  {RGB}{174, 199, 232}
\definecolor{Torange}{RGB}{255, 187, 120}
\definecolor{Tgray}  {RGB}{199, 199, 199}
\definecolor{Tpurple}{RGB}{197, 176, 213}
\definecolor{Tpink}  {RGB}{247, 182, 210}

\title{ArborX: A Performance Portable \UP{Geometric} Search Library}

\ifjournal
  \acmJournal{TOMS}


  \author{D.~Lebrun-Grandi\'e}
  \orcid{0000-0003-1952-7219}
  \affiliation{
    \institution{Oak Ridge National Laboratory}
  }
  \email{lebrungrandt@ornl.gov}
  \author{A.~Prokopenko}
  \orcid{0000-0003-3616-5504}
  \affiliation{
    \institution{Oak Ridge National Laboratory}
  }
  \email{prokopenkoav@ornl.gov}
  \author{B.~Turcksin}
  \orcid{0000-0001-5954-6313}
  \affiliation{
    \institution{Oak Ridge National Laboratory}
  }
  \email{turcksinbr@ornl.gov}
  \author{S.~R.~Slattery}
  \orcid{0000-0003-0103-888X}
  \affiliation{
    \institution{Oak Ridge National Laboratory}
  }
  \email{slatterysr@ornl.gov}

  \date{}
\else
  \author{D.~Lebrun-Grandi{\'e}\thanks{Oak Ridge National Laboratory (\texttt{lebrungrandt@ornl.gov})},\;
          A.~Prokopenko\thanks{Oak Ridge National Laboratory (\texttt{prokopenkoav@ornl.gov})},\;
          B.~Turcksin\thanks{Oak Ridge National Laboratory (\texttt{turcksinbr@ornl.gov})},\;
          S.~R.~Slattery\thanks{Oak Ridge National Laboratory (\texttt{slatterysr@ornl.gov})}}

  \date{}
\fi

\begin{document}

\ifjournal
\else 

\maketitle

\begin{abstract}
  Searching for geometric objects that are close in space is a fundamental
  component of many applications. The performance of search algorithms comes
  to the forefront as the size of a problem increases both in terms of total
  object count as well as in the total number of search queries
  performed. Scientific applications requiring modern leadership-class
  supercomputers also pose an additional requirement of performance
  portability, i.e.~being able to efficiently utilize a variety of hardware
  architectures.  In this paper, we introduce a new open-source C++ search
  library, ArborX, which we have designed for modern supercomputing
  architectures.  We examine scalable search algorithms with a focus on
  performance, including a highly efficient parallel bounding volume hierarchy
  implementation, and propose a flexible interface making it easy to integrate
  with existing applications. We demonstrate the performance portability of
  ArborX on multi-core CPUs and GPUs, and compare it to the state-of-the-art
  libraries such as Boost.Geometry.Index and nanoflann.
\end{abstract}

\fi 

\blfootnote {%
This manuscript has been authored by UT-Battelle, LLC, under contract
DE-AC05-00OR22725 with the U.S. Department of Energy. The United States
Government retains and the publisher, by accepting the article for
publication, acknowledges that the United States Government retains a
nonexclusive, paid-up, irrevocable, world-wide license to publish or reproduce
the published form of this manuscript, or allow others to do so, for United
States Government purposes.
}

\section{Introduction}\label{sec:intro}

Performing proximity searches on collections of geometric objects is an
inherent component of applications in many fields. Finding the nearest
neighbors of a point, or finding all objects within a certain distance, are
common tasks in shape registration methods~\cite{besl1992} in computer vision
and pattern recognition, special effects in games and
movies~\cite{karras2012a}, anomaly detection~\cite{breunig2000lof}, machine
learning~\cite{scikit-learn}, \UP{cosmology~\cite{halo2015}}, data transfer in multiphysics
simulations~\cite{slattery2016}, contact detection in computational
mechanics~\cite{feng2002augmented}, and many others. Such algorithms involve
multiple searches through thousands or millions of objects. The performance of
search algorithms is thus crucial for the overall performance of an
application.  Brute force computations are prohibitively expensive for all but
the simplest applications with very few objects of interest. Instead, methods
employing tree-based data structures are preferred due to their inherent
logarithmic cost.

Many libraries have been developed dedicated to search algorithms. A major
choice in developing such a library is the underlying tree data
structure. This choice dictates both the complexity of implementation and the
resulting performance, including the tradeoff between the time to construct
the data structure and the time to perform search queries. A data structure
that is fast in performing the search may require a longer time setting up,
and vice versa. The way it is used in each application dictates the desired
tradeoff. Two tree structures, k-d and R-tree, are particularly suitable for
geometry-based search, and are commonly implemented in libraries.

k-d tree~\cite{bentley1975} is a binary space partitioning data structure. The
construction algorithm chooses a suitable hyperplane to split a given set of
points into two, and continues recursively for each subset, deciding on a new
hyperplane each time. The internal nodes of the tree correspond to such
hyperplanes, with the parent-child relationship formed through a single recursive
iteration. The hyperplane orientations are typically switched at each level so as
not to produce very skewed sets, with cyclic rotation amongst dimensions being
the simplest approach. Once the algorithm terminates, the leaf nodes contain
the original set of points. Variants of the k-d tree are widely used in libraries,
e.g., FLANN~\cite{muja2009} and nanoflann~\cite{blanco2014}.

The R-tree~\cite{guttman1984} is an alternative data structure used for spatial
search. The leaf nodes of an R-tree are multidimensional rectangles bounding
the objects of interest, and higher level nodes of a tree are aggregations of an
increasing number of objects. This is the data structure that was chosen in the
Boost.Geometry.Index~\cite{boost_geometry} library.

Both nanoflann and Boost.Geometry.Index libraries are widely used in
applications. Both, however, are less suitable for high-performance computing
(HPC) applications as they do not take advantage of multi-threading, nor do
they consider the variety of different architectures available today. Current
HPC trends require search algorithms to perform well on a variety of hardware
architectures, including GPUs and other accelerators provided by a variety of
vendors. This is particularly true within the Exascale Computing Project
of the U.S. Department of Energy~\cite{exascale_computing_project} where
significant resources are devoted to porting applications to utilize both CPUs
and GPUs, and preparing for upcoming new architectures such as APU and
FPGA. Given the variety of the current hardware landscape and some uncertainty
in future hardware directions, a search library that is developed from scratch
should be performance portable.

With this goal in mind, we introduce a new open-source library ArborX\footnote{
\UP{ArborX is available at \url{https://github.com/arborx/ArborX}}.}. It is a
header-only C++ library with a focus on performance portability for both current
and known future leadership-class supercomputers. The implemented algorithms
were carefully chosen to be efficient on the multiple architectures, and rely
on the C++ Kokkos~\cite{carter2014} library to provide performance
portability. We focus on low order dimensional space, building the data
structures from scratch (i.e., no incremental updates).

The paper is organized as follows. In~\Cref{sec:search}, we describe the
algorithms that are used in ArborX. In~\Cref{sec:numerical_results}, we
compare our library to other state-of-the-art libraries and demonstrate its
performance on different architectures. Finally, we present our conclusions
in~\Cref{sec:conclusions}.

\section{Search algorithm}\label{sec:search}

Search algorithms are memory bound by nature. The fundamental parts of any
good search algorithm include visiting as few tree nodes of the search tree as
possible, reducing the amount of memory required by each tree node, and using
inexpensive computations to construct and query the tree data
structure. Furthermore, reducing thread execution divergence (executing
different code) and data divergence (reading or writing disparate locations in
memory) is highly desirable in parallel implementations, particularly for
accelerators with thousands of threads (such as GPUs) and architectures that
improve performance via vectorization (e.g. modern CPUs). Below we present a
bounding volume hierarchy (BVH) tree data structure that was carefully chosen
to satisfy all of these requirements on modern architectures.

A BVH is a tree structure created from a set of geometric objects in a
multi-dimensional space. The objects are wrapped in simple geometric form
(\textit{bounding volumes}) that form leaf nodes of the tree. Similar to the
R-tree, each node of a BVH is an aggregate of its children, enclosing the
group within a larger bounding volume. The root node of the hierarchy
corresponds to the bounding volume around all objects (called scene bounding
volume). Binary BVH is by far the most popular choice and is what we have
chosen for our implementation in this work. For multithreaded and GPU
implementations the binary BVH has the convenient property that the number of
internal nodes in the tree is equal to the number of leaf nodes decreased by
one which allows for static memory allocations once the input geometry is
known.

\begin{figure}[t]
  \centering
  \begin{subfigure}[t]{0.45\textwidth}
    \centering
    \definecolor{mycolor}{HTML}{F14B2E}

\tikzset{
  internal_zero/.style = {thick,color=black},
  internal_one/.style = {thick,color=black},
  internal_two/.style = {thick,color=black},
  internal_three/.style = {thick,color=black},
  leaf/.style = {thick,color=mycolor},
}

\tiny
\begin{tikzpicture}[scale=2]

\draw[internal_zero] (0,0.1) rectangle (1.04,1.08);
\node[anchor=north west] at (0,1.08) {14};
\draw[internal_one] (0.5,0.2) rectangle (1.04,1.08);
\draw[internal_two] (0.6,0.7) rectangle (0.94,1.08) node[anchor=north east] {11};
\node[anchor=south] at (0.77,0.58) {13};
\draw[leaf] (0.8,0.7) rectangle node {\small \faFemale} node[anchor=center,color=black] {7} (0.94,0.88);
\draw[leaf] (0.6,0.9) rectangle node {\small \faMale} node[anchor=center,color=black] {2} (0.74,1.08);
\draw[internal_two] (0.5,0.2) rectangle (1.04,0.58) node[anchor=north east] {10};
\draw[leaf] (0.9,0.25) rectangle node {\small \faFemale} node[anchor=center,color=black] {0} (1.04,0.43);
\draw[internal_three] (0.5,0.2) rectangle (0.84,0.58);
\node[anchor=south west] at (0.5, 0.2) {9};
\draw[leaf] (0.5,0.4) rectangle node {\small \faFemale} node[anchor=center,color=black] {6} (0.64,0.58);
\draw[leaf] (0.7,0.2) rectangle node {\small \faMale} node[anchor=center,color=black] {4} (0.84,0.38);
\draw[internal_one] (0,0.1) rectangle (0.34,0.88);
\node[anchor=north west] at (0,0.88) {12};
\draw[leaf] (0.2,0.7) rectangle node {\small \faFemale} node[anchor=center,color=black] {3} (0.34,0.88);
\draw[internal_two] (0,0.1) rectangle (0.30,0.38);
\node[anchor=north west] at (0,0.40) {8};
\draw[leaf] (0.16,0.2) rectangle node {\small \faFemale} node[anchor=center,color=black] {1} (0.30,0.38);
\draw[leaf] (0,0.1) rectangle node {\small \faMale} node[anchor=center,color=black] {5} (0.14,0.28);


\end{tikzpicture}
    \caption{Geometric objects and bounding volumes\label{f:bvh_human:volumes}}
  \end{subfigure}
  \begin{subfigure}[t]{0.45\textwidth}
    \centering
    \includegraphics[width=0.8\textwidth]{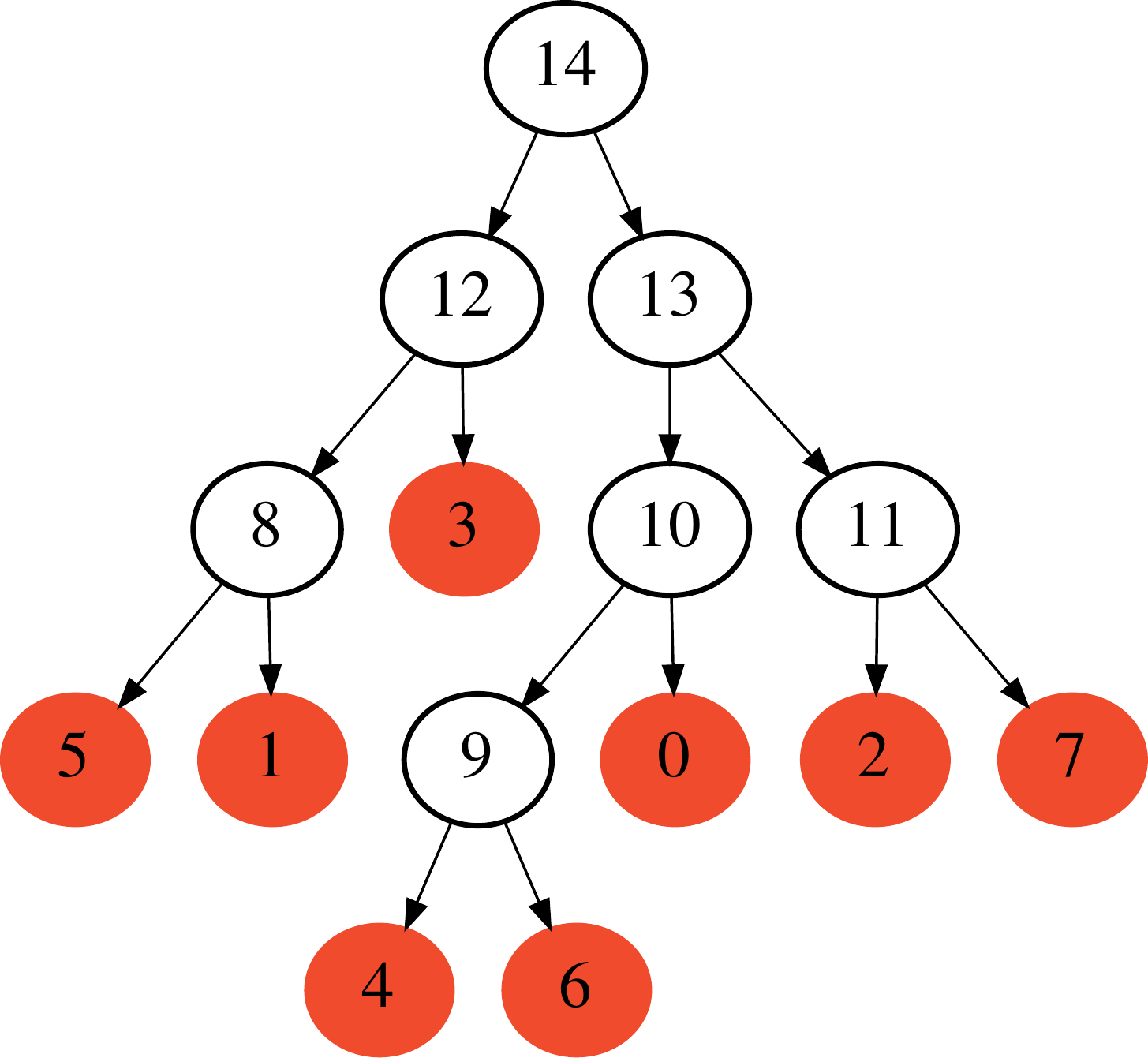}
    \caption{BVH tree\label{f:bvh_human:bvh}}
  \end{subfigure}
  \caption{Bounding volumes and corresponding BVH tree for a set of eight geometric
  objects represented by human figures.}
  \label{f:bvh_human}
\end{figure}
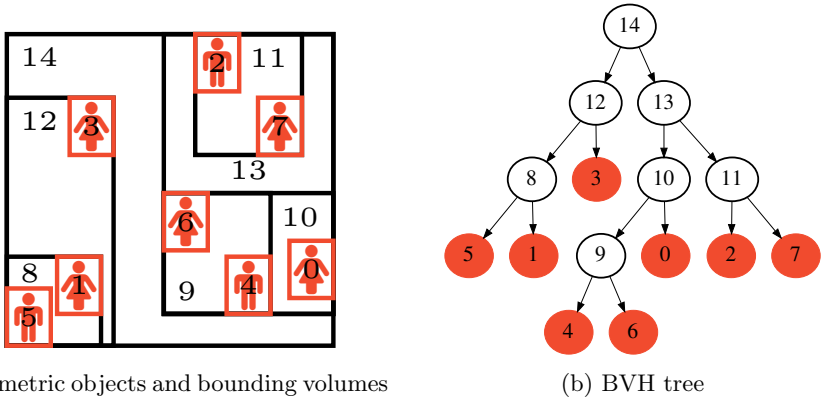

The choice of the geometry of a bounding volume is crucial for performance.
Bounding volumes should require little data to store, be fast to test for
intersection, have fast distance computations, and fit closely to the
underlying object (to avoid unnecessary traversal). In practice, axis-aligned
bounding boxes (AABB), which are boxes aligned with axes of the coordinate
system, are often a good choice~\cite{haverkort2004}. They require minimal
space to store (two opposite corner points, or six \UP{floating point} numbers in
3D) and are fast to test for intersections. Computing the distance from a
point to an AABB is also inexpensive. This often outweighs the main drawback
of AABB of not fitting tightly to the underlying data in some
situations. \Cref{f:bvh_human} demonstrates an example of a BVH tree formed
for a set of eight geometric objects represented by human figures. The red
bounding volumes correspond to leaf nodes (nodes 0-7) and tightly surround the
objects themselves. They are then combined to form larger bounding volumes
corresponding to internal nodes (nodes 8-14), culminating with the root node
bounding volume surrounding the whole scene. The corresponding BVH tree is
shown in~\Cref{f:bvh_human:bvh}.

\UP{
The idea of parallelization of BVH construction by using a space-filling curve
(called linear BVH) was proposed in~\cite{lauterbach2009fast}. In that
approach, the leaves of a tree are ordered based on a space-filling Z-curve
using Morton codes. The algorithm was improved in~\cite{pantaleoni2010hlbvh}
and~\cite{garanzha2011simpler} by allowing processing each level in parallel.
In~\cite{karras2012d}, a new approach allowing all internal nodes be
constructed concurrently was introduced. In~\cite{apetrei2014}, the algorithm
was further improved by merging hierarchy construction with bounding volume
computations in a single bottom-up pass.

The question of a quality of the constructed hierarchy often arises in
applications. For ray tracing applications, there is a rich literature
devoted to the question (see, for example, \cite{domingues2015} and the
references within). Typically, the goal is to minimize the surface area
heuristic (SAH), which is often achieved through rearranging subtrees. It is
possible that many of the techniques could be applicable to scientific
applications once an applicable heuristic is determined. Other approaches may
warrant examination. In~\cite{vinkler2017}, the authors propose extending
Morton codes to incorporate additional geometric information, such as object
size, leading to higher quality BVH. Uneven primitive distribution is addressed
in \cite{yingsong2019} through computation of a local density information to
inform the BVH construction.

Another active area of research is reducing the amount of bandwidth used
by BVH. \cite{howard2019quantized} replaces single precision floating points
used to store a bounding volume with two 32-bit integers containing quantized
bounds. \cite{benthin2018} reduces the depth of the hierarchy through
compressing leaf nodes into wide multi-nodes.

The focus of this work has so far been on the speed and portability of the
library, rather than the quality of the constructed hierarchy. In scientific
applications, it is typical that the tree is rebuilt multiple times (e.g.,
for each time step of a time-dependent application), placing lower importance
on the quality. We chose~\cite{karras2012d} for our initial implementation
due to its simplicity. Addressing the quality is deferred to the future work.
}

We provide an overview of the core algorithms for the construction of such tree
in~\Cref{sec:bvh_construction}. The corresponding traversal algorithms are
detailed in~\Cref{sec:bvh_traversal}. Finally, in~\Cref{sec:API}, we discuss a
user interface to allow flexibility in interaction with user data.

\subsection{BVH construction}\label{sec:bvh_construction}

In this section, we describe the construction of the linear BVH used to
accelerate the search for a given set of geometric objects. The degree of
parallelism in BVH trees is severely limited in a typical bottom-up
construction (i.e., constructing a node only after its children). On the other
hand, the linearity imposed by a Z-curve implicitly partitions the objects
based on the highest differing bit of their Morton codes, allowing for a
top-down approach. A clever numbering of internal nodes as described
in~\cite{karras2012d} then allows for a fully parallel algorithm. \UP{We
implement the original algorithm with only minor changes (such as
removing parent pointers from tree nodes, and storing the leaf node permutation
index in a leaf), with an intent to incorporate~\cite{apetrei2014} in the near
future.} A brief description of the involved steps is provided below.

\textit{Construct AABBs.} As in any BVH algorithm,
the first step is to compute the bounding boxes of the user provided objects.
The only requirement on the objects is that they are \textit{boundable}. For
certain classes of objects, such as polyhedrons, this step is inexpensive. The
computed boxes may be degenerate, such as those produced for points or objects
of a dimension lower than that of AABBs, giving one or more dimensions with an
extent of zero.

\textit{Calculate the scene bounding box}. The scene bounding box is an AABB
that contains the bounding boxes of all objects. It is easily computed by a
reduction of the corners of the bounding boxes.

\textit{Assign Morton codes for each AABB.} Morton codes, or Z-order codes,
are used to map multidimensional data to a single dimension, while preserving
the spatial locality of the data. Given a point, a Morton code can be
efficiently computed by interleaving bits of the point coordinates. The
Morton code of a bounding box is computed as the Morton code of its centroid
scaled using the scene bounding box. This guarantees that all coordinates lie
within $[0,1]^3$ cube. In general, Morton codes are not guaranteed to be
unique. Thus, if multiple objects share the same Morton code, they are
augmented with an index to differentiate them.

\textit{Sort the bounding boxes using their Morton code.}
The bounding boxes may now serve as leaf nodes.
The goal of this procedure is to decrease the size and the overlap of bounding boxes
of internal nodes that will be generated.

\textit{Generate the bounding volume hierarchy.}
With a linear order imposed by sorted Morton codes, construction of a hierarchy
can be seen as a recursive partitioning of the range of Morton indices so that
each internal node in a tree corresponds to an interval of Morton codes. The
recursion terminates when a range contains only one item, which is to be leaf
node. The described partitioning is based on selecting a position called
\textit{split} to cut a given range in two.
The splits are based on the highest differing bits of Morton codes
within a given range.
The range of each internal node is computed independently, allowing a parent node to
determine its children and record the parent-child relationships without
waiting for the construction of other nodes. A more detailed discussion can be
found in~\cite{karras2012d}. \UP{The permutation indices computed in the previous
step are stored in the leaf nodes.}

\textit{Calculate internal nodes bounding boxes.} The final step is to compute
the bounding boxes of internal nodes by traversing the tree bottom-up. In
parallel, each thread is assigned a leaf node and traverses towards the tree root. Upon
encountering an internal node, only one of the children's threads is allowed to
proceed further. \UP{As parent pointers are not used in hierarchy traversal, we
avoid storing the pointer to a parent node inside a child node to minimize used
memory. Instead, the parent pointers are kept in an auxiliary array that is
dismissed after construction.}

\subsection{BVH traversal}\label{sec:bvh_traversal}

Once constructed the tree data structure may be used as many times as needed
to complete the search process. Each query of the data structure results in a
traversal of the tree where the approximation of objects by AABB boxes allows
for a preliminary coarse search which is responsible for listing all boxes
with potential collisions. It is then followed by a fine search where
a user-specified search criteria is used to trim the results. The tighter the
bounding volumes are to the real objects, the more accurate the results of the
coarse search are, and the fewer expensive fine search queries that are needed.

We distinguish two kinds of (search) queries; spatial and nearest. A spatial
query searches for all objects within a certain distance of an object of
interest. A nearest query, on the other hand, looks for a certain number of
closest objects regardless of their distance from an object of interest. These
two query kinds require fundamentally different tree traversal algorithms. The
spatial query has to necessarily explore all nodes in a tree that satisfy a
given distance-based predicate. The nearest query, on the other hand, can
terminate early when it can be guaranteed that the already found candidates
are the best possible ones.

It is very common to execute multiple search queries simultaneously. In
parallel, the threads are executed in \emph{batched} mode, with each thread
assigned a range of search queries (on CPU) or a single search query (on GPU).
While it may be possible to further improve the performance by having multiple
threads work on the same query, we do not address this in the current ArborX
implementation, and each query is performed exclusively by a single thread.

As threads traverse the tree, the attention to execution and data divergence is
paramount for performance. We next describe our strategies for traversal with
each query type.

\subsubsection{Traversal for spatial queries}

In spatial query traversal, each node can be tested independently for predicate
satisfaction. A simple distance-based predicate tests for whether a distance
from a AABB to a bounding box is less than a given radius.

Spatial traversal is executed top-down, starting from the root node.
A naive recursive implementation may lead to a high execution divergence as shown
in~\cite{karras2012b}. Instead, an iterative traversal is preferred, using a
stack to keep track of nodes to visit. In the beginning of the traversal the
root node is added to the stack. The algorithm proceeds by popping a node from
the stack, and testing its children for predicate satisfaction, upon which they
are either added to the stack (internal nodes) or to the output (leaf nodes).
The algorithm terminates upon an empty stack.

An important issue associated with spatial traversals is that the number of
found objects is not known \emph{a priori}. \UP{This issue is typically not
addressed in computer graphics applications as the results are processed on the
fly in many cases. However, storing the results plays an important role in
scientific applications, where further processing of the results is required
(e.g., halo finding algorithm~\cite{halo2015} calculates clusters based
on the computed data). It is well known that dynamic memory allocation} is
inefficient in multithreading, and is problematic on GPUs. This can be avoided
by using a count-and-fill technique, i.e., by doing two passes (2P). The first
pass just counts the number of found objects. Then, the required storage is
allocated, and the process is repeated in the second pass, this time storing
the results.

2P approach, while robust, comes with a drawback of having to traverse the
hierarchy twice. A superior alternative, when possible, is to only do the
second pass once the preallocated memory is exceeded. In this approach (called 1P),
an estimate for a maximum number of found objects per query is provided by a
user. During the first pass, the found objects are both counted and stored. If
the storage is exceeded, the algorithm falls back to the 2P approach. If the
estimate is correct, i.e., is an upper bound, only a single pass is done,
improving the overall performance. That single pass is then followed by
``compacting'' the results due to excess allocation. Such a technique is
typically less costly than performing the traversal twice.

\subsubsection{Traversal for nearest queries}

Nearest traversal proceeds in top-down fashion, similarly to spatial-based
traversal. However, in this case, the number of found neighbors (or, rather,
its upper bound) is known in advance, and thus allows for the preallocation of
memory and to avoid the second pass through the tree.

A typical implementation of nearest traversal uses a priority queue based on
distances, using the closest node in each iteration. An alternative and better
performing approach, first derived for k-d trees in~\cite{patwary2016}, is to
use a stack.
As stack is a Last-In-First-Out data structure, it is possible to get a
behavior similar to the one of a priority queue by adding a child with a
shorter distance second (so that it sits on top of the stack). The algorithm
terminates when the remaining candidates in the stack are guaranteed to result
in worse results, or the stack is empty. The final (optional) step is to clean
the results by purging missing data (if, for example, the number of found
objects is less than specified).

\subsubsection{Query ordering}\label{sec:bvh_query_sort}

Both execution and data divergence depend heavily on whether the nearby
computational threads traverse the tree similarly. Sorting the queries may have
a significant impact on the overall performance as was noted
in~\cite{karras2012b}. One way to accomplish this is by making sure that the
search queries for nearby threads are ``close'' to each other, i.e. the
corresponding tree traversals would be very similar in both nodes they visit
and the order they visit them. This can be achieved by computing Morton codes
for query objects and then using them for pre-sorting.

\begin{figure}[t]
  \begin{subfigure}[t]{0.32\textwidth}
    \centering
    \includestandalone[width=0.9\textwidth]{figures/ornl_leaf}
    \caption{Point cloud\label{f:query_sort:cloud}}
  \end{subfigure}
  \begin{subfigure}[t]{0.32\textwidth}
    \centering
    \includegraphics[width=\textwidth]{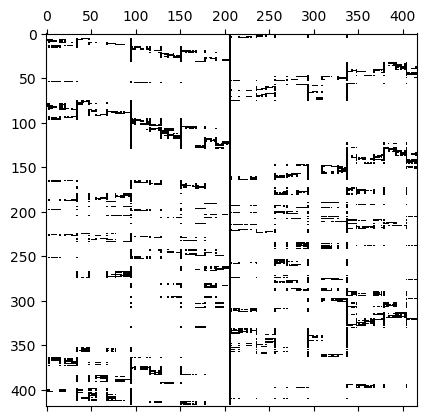}
    \caption{Queries in original order\label{f:query_sort:original}}
  \end{subfigure}
  \begin{subfigure}[t]{0.32\textwidth}
    \centering
    \includegraphics[width=\textwidth]{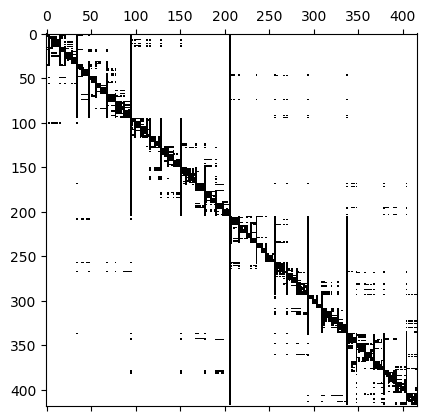}
    \caption{Queries in Morton-based sorted order\label{f:query_sort:sorted}}
  \end{subfigure}
  \caption{Effect of query ordering on nearest traversal.\label{f:query_sort}}
\end{figure}

To illustrate this phenomenon, consider the nearest traversal of a point cloud of
418 points (representing a leaf) shown in~\Cref{f:query_sort:cloud}.
\Cref{f:query_sort:original} represents a binary matrix of size $418 \times
418$ corresponding to the original ordering of search queries. Each row
represents a search query assigned to a single thread, and each column
corresponds to an internal node in the tree. A value in the matrix is nonzero
(black) when a thread accesses a bounding box of a corresponding node. As one
can see, using the original ordering of search queries results in little
correlation of accessed nodes of two nearby threads. The performance suffers in
this case due to poor memory access pattern.
\Cref{f:query_sort:sorted}, on the other hand, corresponds to the queries
reordered based on their Morton codes. It is clear that the nearby threads now
share many nodes of the tree in their traversal. The apparent hierarchical
pattern of the matrix indicates the concentration of queries in certain
subtrees before switching to sibling trees.

\subsection{Library interface}\label{sec:API}

ArborX implements a performance portable interface through the use of
Kokkos~\cite{carter2014}, a C++ library providing a uniform programming
interface for various backends, such as OpenMP or CUDA. Using Kokkos allows for
running the same code on CPUs or GPUs by simply changing the backend through a
template parameter.

\begin{figure}[t]
  \centering
  \begin{minipage}{0.83\textwidth}
  \begin{lstlisting}[language=C++, frame=single, columns=fullflexible]
 // Create the View for the bounding boxes
 Kokkos::View<ArborX::Box*, DeviceType> bounding_boxes("bounding_boxes", n_boxes);
 // Fill in the bounding boxes Kokkos::View
 ...
 // Create the bounding volume hierarchy
 ArborX::BVH<DeviceType> bvh(bounding_boxes);
  \end{lstlisting}
  \end{minipage}
  \caption{BVH construction interface.}
  \label{f:bvh_construction}
\end{figure}


The construction procedure begins with a set of bounding boxes, provided by a
user as a \code{Kokkos::View}, a Kokkos data structure corresponding to a
multi-dimensional array. At a high level, \code{Kokkos::View<T*, DeviceType>}
can be thought of as an array containing objects of type \code{T}. The
\code{DeviceType} template argument indicates both the memory that the data
resides in (e.g., host or device memory), and the place to execute the code
(e.g., CPU or GPU). Once bounding boxes are constructed, they are passed to
the constructor of \code{BVH}, the ArborX class containing the hierarchy (see
\Cref{f:bvh_construction}).

\begin{figure}[t]
  \centering
  \begin{minipage}{0.83\textwidth}
  \begin{lstlisting}[language=C++, frame=single, columns=fullflexible]
 // Create the View for the spatial-based queries
 Kokkos::View<ArborX::Within *, DeviceType> queries("queries", n_queries);
 // Fill in the queries
 using ExecutionSpace = typename DeviceType::execution_space;
 Kokkos::parallel_for("setup_queries",
     Kokkos::RangePolicy<ExecutionSpace>(0, n_queries), KOKKOS_LAMBDA(int i) {
       queries(i) = ArborX::within(query_points(i), radius);
    });
 // Perform the search
 Kokkos::View<int*, DeviceType> offsets("offset", 0);
 Kokkos::View<int*, DeviceType> indices("indices", 0);
 bvh.query(queries, indices, offsets, buffer_size);
  \end{lstlisting}
  \end{minipage}
  \caption{BVH search interface (spatial queries).}
  \label{f:bvh_radius}
\end{figure}


Next, the queries are built. Each query corresponds to a pair of a query point
and a number of neighbors to be found (nearest query), or a pair of a query
point and a radius (spatial query). The spatial-based version is shown
in~\Cref{f:bvh_radius}, with queries being filled in a device parallel loop.
Once the queries are constructed, two views are allocated to store the results:
the~\code{indices} view to contain the indices associated with the bounding
boxes that satisfy the queries, and the~\code{offsets} view to contain the
offsets in~\code{indices} associated with each query. Two views are necessary
as the number of results for each query may differ (for example, the number of
results within specific radius for spatial-based search)\footnote{This format
is similar to that of compressed sparse row format that is commonly used to
store sparse matrices.}. The search is done by invoking the \code{query}
function of BVH. The additional parameter \code{buffer\_size} is optional, and
only used for spatial-based queries. It indicates the user-provided estimate
for the number of returned results, and if accurate, allows to do a single pass
(1P).

\section{Numerical Results} \label{sec:numerical_results}

The numerical studies presented in the paper were performed on two systems:
\begin{itemize}
 \item CADES system with each node containing two Intel Xeon E5-2695 v4 18-core
   CPUs running at a clock speed of 2.1 GHz with 256 GB of main memory;
\item OLCF Summit system with each node having two IBM POWER9 AC922 21-core
  CPUs, each having 4 hardware threads with 6 Nvidia Volta V100 GPUs connected
    by NVLink 2.0~\cite{olcf_summit}.
\end{itemize}

We used the Google Benchmark tool~\cite{google_benchmark} in our experiments, using
the median of the runs for the results we have reported here.

\subsection{Experimental Data Sets}

In our experiments, we use several artificial data sets proposed
in~\cite{elseberg2012}. We consider two shape forms, cube and sphere. For a
given shape, a set of points is then chosen either from within the selected
shape (filled variant), or from its boundary (hollow variant). To generate $p$
points, set $a = p^{1/3}$, $\Omega = [-a,a]^3$, and proceed as follows:
\begin{itemize}
  \item
    \textit{filled cube:} each random point is drawn randomly from $\Omega$ with uniform
    distribution;
  \item
    \textit{hollow cube:} points are placed on the faces of $\Omega$ in a cyclic manner,
    with the position of the point on each face being random with uniform
    distribution;
  \item
    \textit{filled sphere:} points are randomly chosen from $\Omega$ and accepted based
    on being within a sphere of radius $a$ centered at 0;
  \item
    \textit{hollow sphere:} points are first generated within $[-1,1]^3$ cube and then
    projected to the sphere of radius $a$ centered at 0.
\end{itemize}

In our experiments, we consider two cases: searching for a filled sphere
cloud of query points in the filled cube cloud (filled case), and searching for
a hollow sphere cloud in the hollow cube cloud (hollow case). The major
difference between these two cases is the workload per thread. For the filled
case, the data and the query results are balanced between threads. The hollow
case, however, presents a challenge due to a wide imbalance of query results, as
only a small number of threads will produce positive results for a spatial search.

For a given problem, $m$ source points and $n$ target points are generated
using one of the described four shapes. The number of neighbors $k$ for the nearest
search is fixed to 10 in all experiments. The radius~$r$ for spatial search is
chosen in such a way that on average there are $k$ neighbors within radius $r$
in a filled cube shape.

\subsection{Comparison with available libraries}

In this section, we compare the performance of ArborX with that of
two state-of-the-art existing libraries, Boost.Geometry.Index and nanoflann.

The Boost.Geometry.Index~\cite{boost_geometry} library implements different
algorithms for R-trees. For the purpose of performance comparison, we used the
packing algorithm~\cite{leutenegger1997,garcia1998} which is the most
performant algorithm contained in Boost.Geometry.Index. The performance comes
at the cost of flexibility since the tree has to be built statically. We used
version~1.67.0 of the library.

nanoflann~\cite{blanco2014} is a header-only library for building k-d trees.
We used nanoflann hash~3b2065e.

As Boost.Geometry.Index and nanoflann are implemented only in
serial\footnote{As Boost.Geometry.Index is thread safe, it is theoretically
possible to run it in batched mode. However, this will require a user to write
the necessary parallel implementation.}, the comparisons in this subsection
were done using one thread. The scaling of ArborX with the number of OpenMP
threads is demonstrated in the next section.

The experiments were performed on the CADES system. They were run
for the increasing number of source points $m$, ranging from $10^4$ to $10^7$.
The number of the target points $n$ was chosen to be the same as the number of
source points, $n = m$. Such configuration is common in many applications,
e.g., finding potentially colliding pairs of objects in graphics applications,
or finding nearby particles for pair-wise interactions in physics simulations.

\begin{figure}[t]
  \begin{center}
    \begin{subfigure}{0.32\textwidth}
      \includegraphics[scale=0.3]{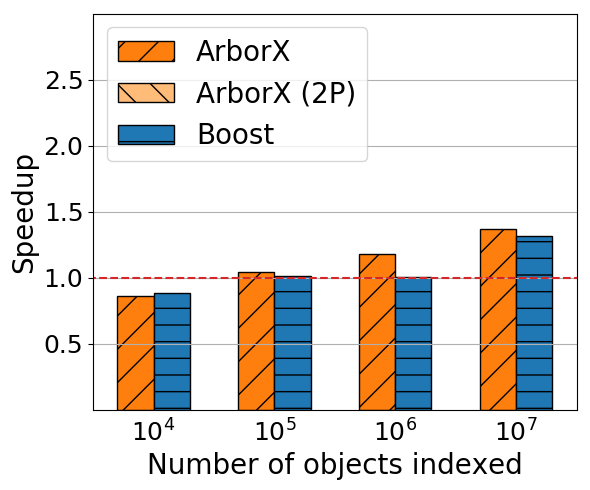}
      \caption{Construction\label{f:comparison_speedup_filled:construction}}
    \end{subfigure}
    \begin{subfigure}{0.32\textwidth}
      \includegraphics[scale=0.3]{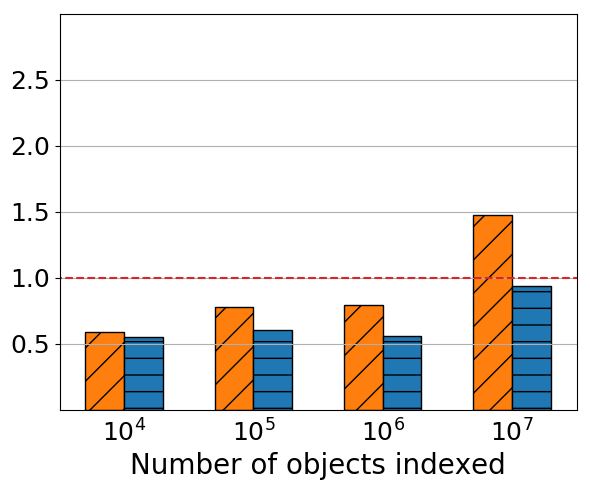}
      \caption{Nearest search\label{f:comparison_speedup_filled:kNN}}
    \end{subfigure}
    \begin{subfigure}{0.32\textwidth}
      \includegraphics[scale=0.3]{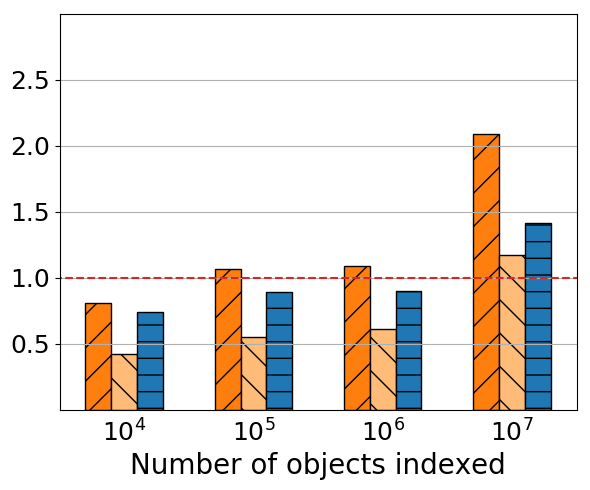}
      \caption{Spatial search\label{f:comparison_speedup_filled:radius}}
    \end{subfigure}
  \end{center}
  \caption{Comparison of libraries for the filled case. The speedup is shown with respect to nanoflann.\label{f:comparison_speedup_filled}}
\end{figure}

\begin{figure}[t]
  \begin{center}
    \begin{subfigure}{0.32\textwidth}
      \includegraphics[scale=0.3]{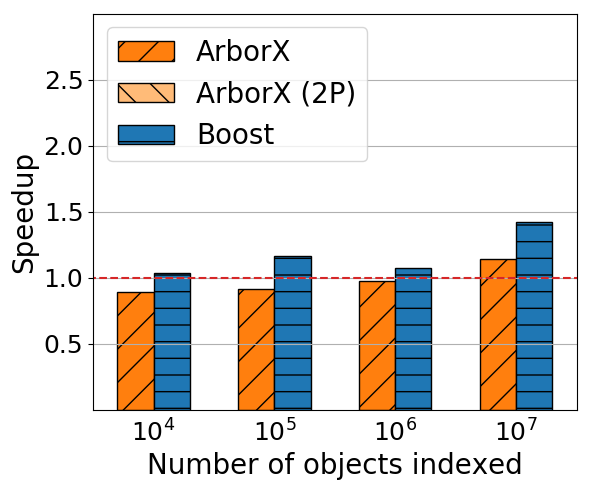}
      \caption{Construction\label{f:comparison_speedup_hollow:construction}}
    \end{subfigure}
    \begin{subfigure}{0.32\textwidth}
      \includegraphics[scale=0.3]{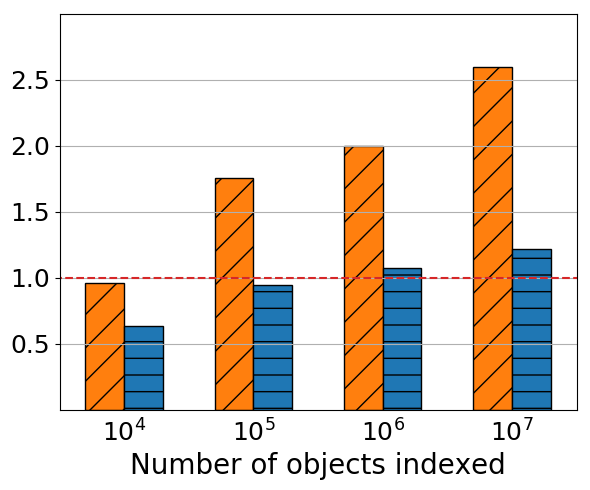}
      \caption{Nearest search\label{f:comparison_speedup_hollow:kNN}}
    \end{subfigure}
    \begin{subfigure}{0.32\textwidth}
      \includegraphics[scale=0.3]{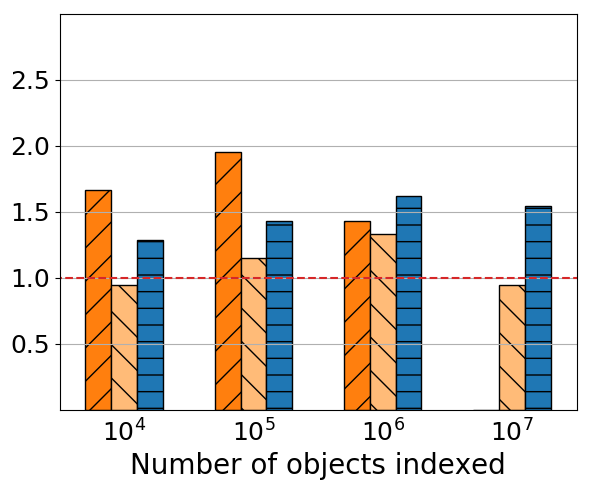}
      \caption{Spatial search\label{f:comparison_speedup_hollow:radius}}
    \end{subfigure}
  \end{center}
  \caption{Comparison of libraries for the hollow case. The speedup is shown with respect to nanoflann.\label{f:comparison_speedup_hollow}}
\end{figure}

\begin{figure}[t]
  \begin{center}
    \begin{subfigure}{0.32\textwidth}
      \includegraphics[scale=0.3]{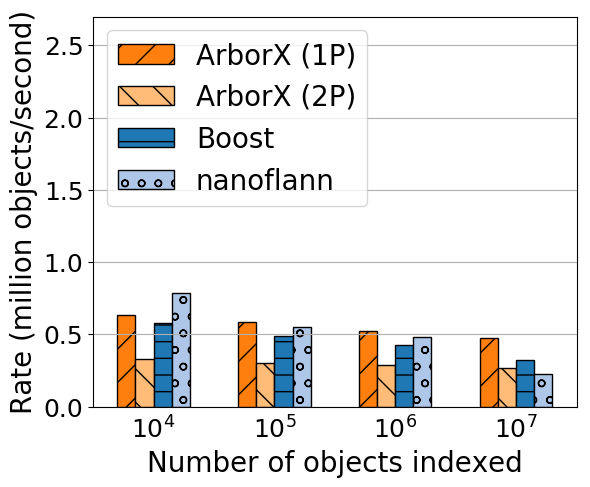}
      \caption{Filled case\label{f:comparison_rate_filled:radius}}
    \end{subfigure}
    \;
    \begin{subfigure}{0.32\textwidth}
      \includegraphics[scale=0.3]{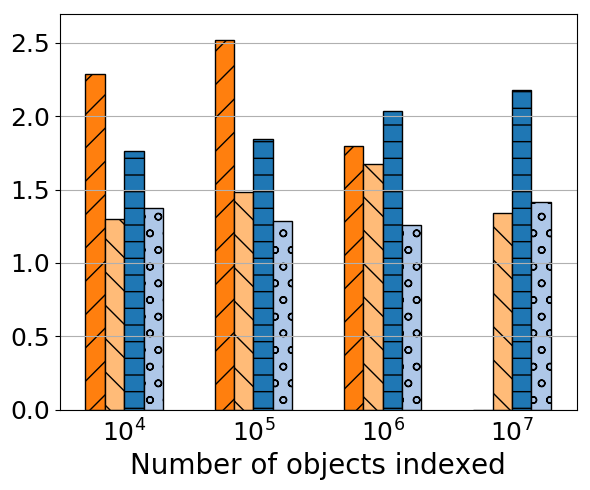}
      \caption{Hollow case\label{f:comparison_rate_hollow:radius}}
    \end{subfigure}
  \end{center}
  \caption{Spatial search rates for the libraries.\label{f:comparison_rate}}
\end{figure}

\Cref{f:comparison_speedup_filled,f:comparison_speedup_hollow}
\UP{demonstrate the relative speedup or slowdown of the libraries relative to
nanoflann}.
\Cref{f:comparison_speedup_filled:construction}~(\ref{f:comparison_speedup_hollow:construction})
shows tree construction speedup for the filled (hollow) case, respectively. We
observe that ArborX and Boost.Geometry.Index libraries perform similarly, while
\UP{nanoflann starts to lose its competitiveness for large number of objects}.
Comparing the query performance of the libraries, we observe that for
the nearest search (\Cref{f:comparison_speedup_filled:kNN,f:comparison_speedup_hollow:kNN}),
ArborX significantly outperforms both
Boost.Geometry.Index and nanoflann libraries for larger numbers of objects,
particularly for the hollow case. For the spatial search, the performance of the 1P
variant is about twice faster than that of the 2P variant
in the filled case (\Cref{f:comparison_speedup_filled:radius}, where
workloads are balanced. However, for a larger number
of objects (larger than $10^6$) in the hollow case, the 1P variant could not be
run due to requiring too much memory to preallocate the storage for the results
based on the maximum estimate. Therefore, no results from the 1P variant are
given for these larger cases in \Cref{f:comparison_speedup_hollow:radius}.

\Cref{f:comparison_rate_filled:radius}~(\ref{f:comparison_rate_hollow:radius})
demonstrate the rate of spatial-based search for a filled (hollow) case,
respectively. The main difference between the filled and hollow variants is the
number of results returned by queries. Specifically, for the spatial-based search,
the filled variant returns 10 neighbors on average (with the minimum being 0 and
the maximum being 32). However, for the hollow variant the number of neighbors is
much more imbalanced, ranging from 0 to 522, with the average being 2. This is due
to a) the fact that the hollow sphere touches the hollow cube in just a few places
(centers of the faces), and b) both being two-dimensional objects thus having a
significantly higher density than in 3D for the same number of points. As
expected, the rate for the hollow variant is significantly faster than that of
the filled variant due to most queries returning empty result.

\UP{An acute observer may notice that there is little difference for between 1P
and 2P variants for the hollow case for large number of objects. It turns out
that the extra memory required to store all the results of the first pass
becomes a drawback at some point, with filtering out unused entries taking a
significant portion of time. Another interesting observation is the drop of
rate for 2P spatial search for the hollow case going from $10^6$ to $10^7$.
Examination revealed that for the latter problem sorting queries is less
effective than retaining their original order. Sorting the queries in serial
may not be necessary, and ArborX provides an option to disable that.}

\subsection{Multi-threaded strong scaling}

We next examine the scalability of ArborX using OpenMP on the CADES system.

For the strong scaling, the number of source points $m$ is fixed to a value
from $10^4$ to $10^7$, and the number of OpenMP threads increased from 1
to 16. The number of target points $n$ was chosen to be the same as the number
of source points, $n=m$.

\begin{figure}[t]
  \begin{center}
    \begin{subfigure}{0.32\textwidth}
      \includegraphics[width=\textwidth]{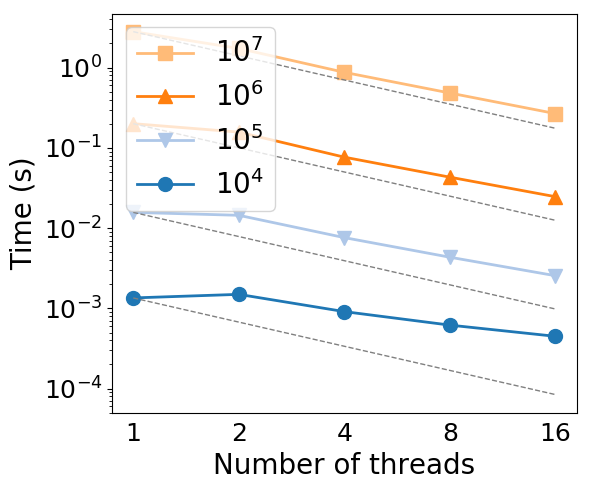}
      \caption{Setup}
    \end{subfigure}
    \begin{subfigure}{0.32\textwidth}
      \includegraphics[width=\textwidth]{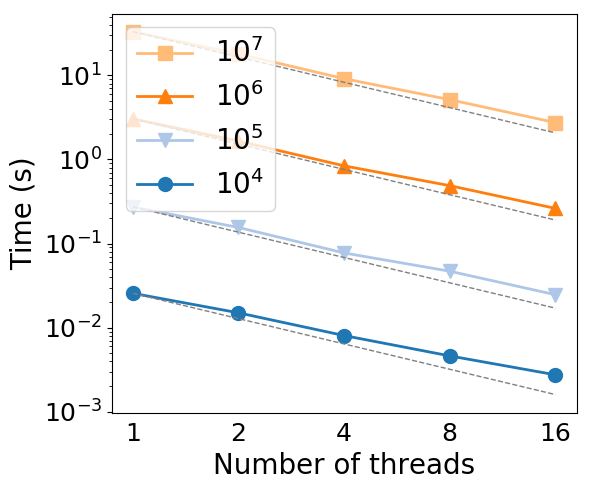}
      \caption{Nearest}
    \end{subfigure}
    \begin{subfigure}{0.32\textwidth}
      \includegraphics[width=\textwidth]{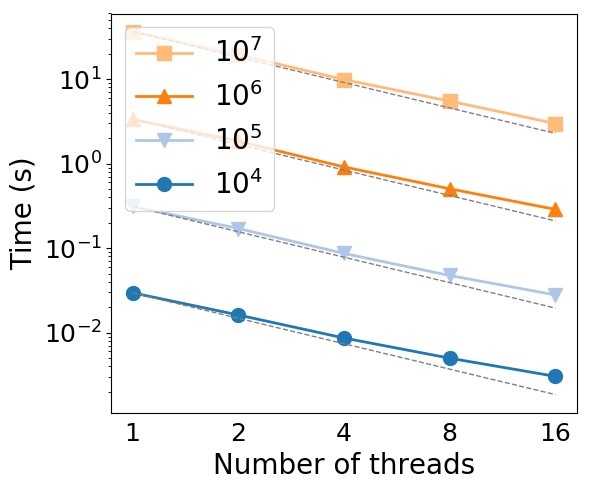}
      \caption{Spatial (2P)}
    \end{subfigure}
  \end{center}
  \caption{ArborX scaling for filled case.\label{f:openmp_scaling_filled}}
\end{figure}

\begin{figure}[t]
  \begin{center}
    \begin{subfigure}{0.32\textwidth}
      \includegraphics[width=\textwidth]{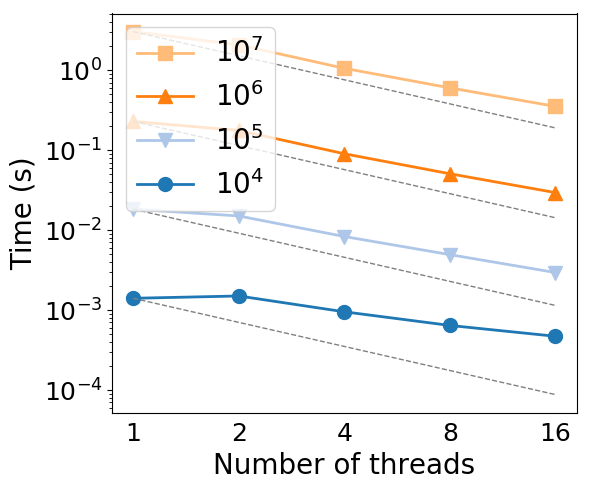}
      \caption{Setup}
    \end{subfigure}
    \begin{subfigure}{0.32\textwidth}
      \includegraphics[width=\textwidth]{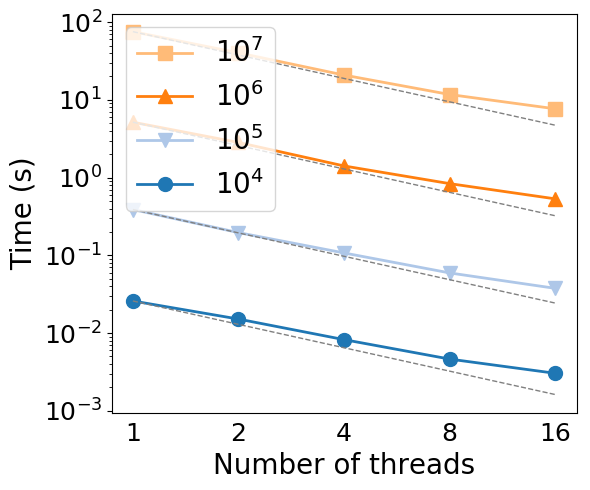}
      \caption{Nearest}
    \end{subfigure}
    \begin{subfigure}{0.32\textwidth}
      \includegraphics[width=\textwidth]{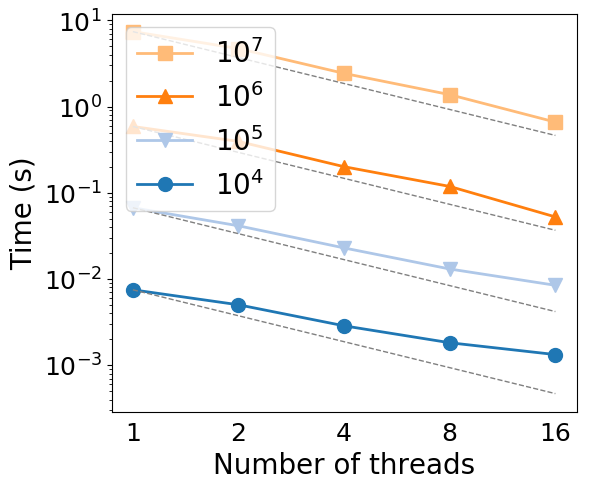}
      \caption{Spatial (2P)}
    \end{subfigure}
  \end{center}
  \caption{ArborX scaling for hollow case. \label{f:openmp_scaling_hollow}}
\end{figure}

The results are presented
in~\Cref{f:openmp_scaling_filled,f:openmp_scaling_hollow}
and~\Cref{t:openmp_scaling_filled,t:openmp_scaling_hollow}. ArborX demonstrates
good scalability for a large number of objects. However, having too few
objects per thread for smaller simulations results in suboptimal
scaling. \UP{Upon further inspection, the sorting routine used for sorting
Morton indices was identified to be the limiting factor, having poor
scalability in cases where every thread has fewer than 1000 objects. We
attempted to use few available parallel sort algorithms (e.g.,
\code{\_\_gnu\_parallel::sort}) instead of the default Kokkos sort. However, we
found this led to only a minor improvement in our performance results.} This
issue affects both construction (sorting of Morton codes,
see~\Cref{sec:bvh_construction}), and search (sorting of queries,
see~\Cref{sec:bvh_query_sort}) \UP{and will be a topic of future work}.

\begin{table}
\caption{ArborX scaling results for the filled case for $n = 10^4$ and $n = 10^7$.\label{t:openmp_scaling_filled}}
\begin{tabular}{ccccccc}
\toprule
\multirow{2}*{Threads} & \multicolumn{2}{c}{Construction} & \multicolumn{2}{c}{Spatial search} & \multicolumn{2}{c}{Nearest search} \\
\cmidrule(lr){2-3} \cmidrule(lr){4-5} \cmidrule(lr){6-7}
& $n = 10^4$& $n = 10^7$& $n = 10^4$& $n = 10^7$& $n = 10^4$& $n = 10^7$\\
\midrule
  1 &  1.00 &  1.00 &  1.00 &  1.00 &  1.00 &  1.00 \\
  2 &  0.90 &  1.62 &  1.83 &  1.87 &  1.71 &  1.84 \\
  4 &  1.48 &  3.21 &  3.42 &  3.69 &  3.18 &  3.62 \\
  8 &  2.18 &  5.82 &  5.93 &  6.64 &  5.55 &  6.44 \\
 16 &  3.01 & 10.50 &  9.74 & 12.24 &  9.31 & 12.06 \\
\bottomrule
\end{tabular}
\end{table}

\begin{table}
\caption{ArborX scaling results for the hollow case for $n = 10^4$ and $n = 10^7$.\label{t:openmp_scaling_hollow}}
\begin{tabular}{ccccccc}
\toprule
\multirow{2}*{Threads} & \multicolumn{2}{c}{Construction} & \multicolumn{2}{c}{Spatial search} & \multicolumn{2}{c}{Nearest search} \\
\cmidrule(lr){2-3} \cmidrule(lr){4-5} \cmidrule(lr){6-7}
 & $n = 10^4$& $n = 10^7$& $n = 10^4$& $n = 10^7$& $n = 10^4$& $n = 10^7$\\
\midrule
  1 &  1.00 &  1.00 &  1.00 &  1.00 &  1.00 &  1.00 \\
  2 &  0.94 &  1.46 &  1.49 &  1.57 &  1.71 &  1.87 \\
  4 &  1.48 &  2.87 &  2.62 &  3.05 &  3.14 &  3.62 \\
  8 &  2.19 &  5.05 &  4.11 &  5.39 &  5.59 &  6.44 \\
 16 &  2.99 &  8.61 &  5.64 & 11.20 &  8.52 &  9.84 \\
\bottomrule
\end{tabular}
\end{table}

\subsection{Accelerator comparison}
We now compare the performance of ArborX using OpenMP and CUDA on the OLCF
Summit system. We compare the performance of the two \UP{POWER9} CPUs on a single
Summit node (42 physical cores) with that of a single Volta V100
GPU\footnote{Currently, Kokkos (and thus ArborX) does not support multiple
  GPUs within the same process. \UP{The six V100 GPUs on a Summit node are typically
  used by having a single MPI rank manage a single dedicated GPU.}}.
POWER9's physical cores consist of four ``slices'' which can be used in a variety
of configurations.
In \emph{smt4} mode, each slice operates independently of the other three,
allowing for separate streams of execution for multiple OpenMP threads on each physical
core.
In \emph{smt2} mode, pairs of slices work together to run tasks.
Finally, in \emph{smt1} mode the four slices work together to execute the task/thread
assigned to the physical core.

\begin{figure}[t]
  \begin{center}
    \begin{subfigure}{0.32\textwidth}
      \includegraphics[width=\textwidth]{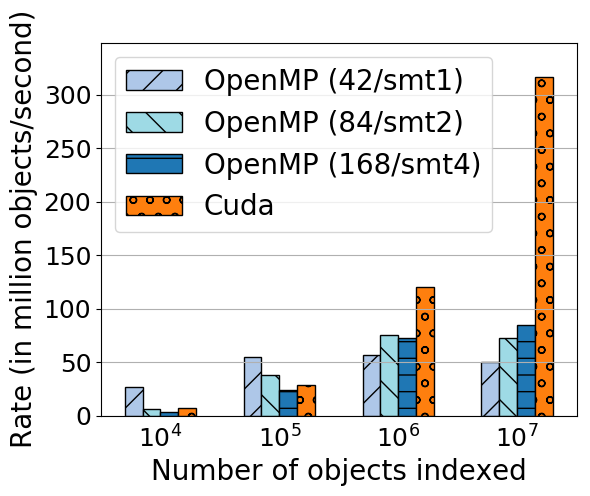}
      \caption{Construction}
    \end{subfigure}
    \begin{subfigure}{0.32\textwidth}
      \includegraphics[width=\textwidth]{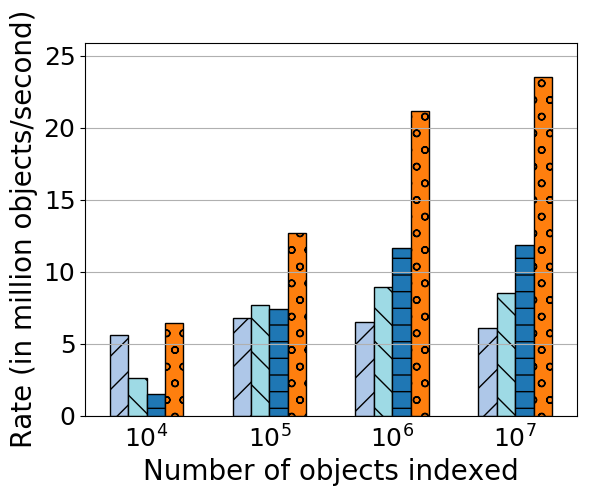}
      \caption{Nearest search}
    \end{subfigure}
    \begin{subfigure}{0.32\textwidth}
      \includegraphics[width=\textwidth]{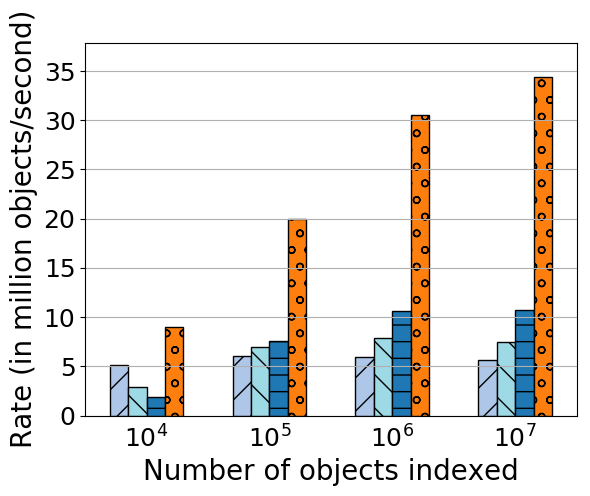}
      \caption{Spatial search}
    \end{subfigure}
  \end{center}
  \caption{Comparison of OpenMP and CUDA on Summit for filled cube source and filled sphere target clouds.}
  \label{f:openmp_cuda_filled}
\end{figure}

\begin{figure}[t]
  \begin{center}
    \begin{subfigure}{0.32\textwidth}
      \includegraphics[width=\textwidth]{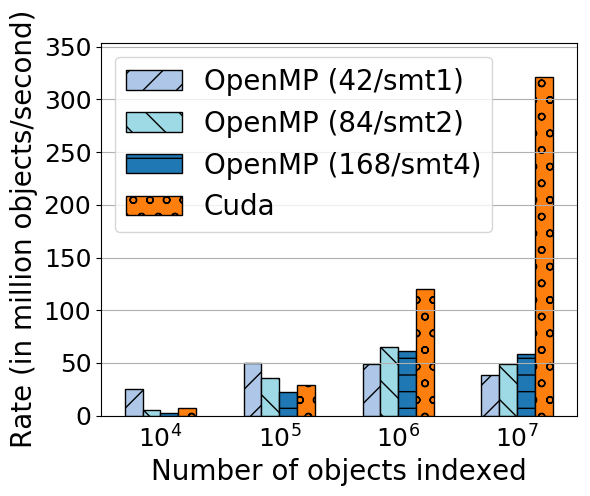}
      \caption{Construction}
    \end{subfigure}
    \begin{subfigure}{0.32\textwidth}
      \includegraphics[width=\textwidth]{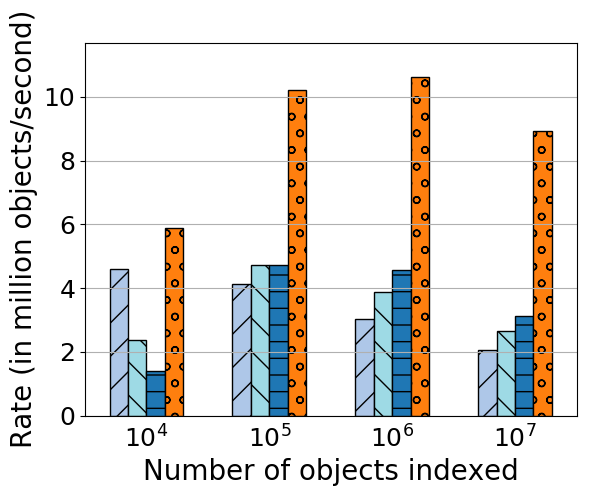}
      \caption{Nearest search}
    \end{subfigure}
    \begin{subfigure}{0.32\textwidth}
      \includegraphics[width=\textwidth]{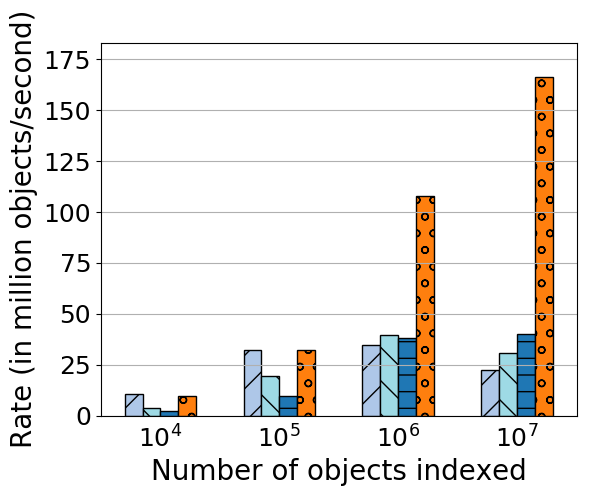}
      \caption{Spatial search}
    \end{subfigure}
  \end{center}
  \caption{Comparison of OpenMP and CUDA on Summit for hollow cube source and hollow sphere target clouds.}
  \label{f:openmp_cuda_hollow}
\end{figure}

\UP{
The results are presented in~\Cref{f:openmp_cuda_filled,f:openmp_cuda_hollow}.
We observe that executing in \emph{smt4} mode usually leads to better
performance (especially, for larger problem sizes), though that is not always
the case. We also note that a single Summit GPU significantly exceeds the
performance of the full node of CPUs, exhibiting shortcomings only for smaller
problems that are less suitable for the high parallelism provided by the
accelerator. As each node of Summit has six GPUs connected by NVLink, it is
expected that using only accelerators would dramatically outperform using only
CPUs when using an MPI+Kokkos approach in the future.
}

We also concede that another way to compare the performance would have been to
normalize the data by the power drawn by the corresponding hardware.
Unfortunately, we were not able to obtain such data as it is not yet provided
by the facility or the performance monitoring software. While the published
specifications of V100 and POWER9 do have some information, including thermal
design power, those were observed to be inaccurate in practice by others, and
thus were not relied on in our reporting.

\section{Conclusion and Outlook}\label{sec:conclusions}

In this paper, we presented a new library ArborX for searching close geometric
objects in space. ArborX's strength lies in its performance and its ability to
be run on multiple hardware architectures using a single interface
definition. Experiments were conducted to compare its performance with
existing popular libraries, such as nanoflann and Boost.Geometry.Index, and to
demonstrate its scalability and performance on accelerators. Our results show
that our implementation is competitive with these libraries in single thread
execution and is also able to effectively leverage both the multithreaded CPU
and GPU compute power on modern leadership-class supercomputers such as
Summit.

There are two natural directions for future work. One is addressing the
current scalability limitations through careful analysis and profiling of the
library.  The second is implementing the distributed search algorithms using
MPI to address the requirements of exascale applications where the objects
indexed by the tree as well as the query objects are distributed across a
large number of MPI ranks. This creates additional challenges to those
presented in this paper as it is likely that the data that one searches for
may not belong to the same node, or that the data distribution among MPI ranks
may be imbalanced. Thus, a communication layer deploying a load balancing
strategy will be required to be effectively scale to thousands of accelerated
compute nodes.

\ifjournal
\begin{acks}
\else
  \section*{Acknowledgements}
\fi
Research sponsored by the Laboratory Directed Research and Development Program
of Oak Ridge National Laboratory, managed by UT-Battelle, LLC, for the U.S.
Department of Energy.

This research was supported by the Exascale Computing Project (17-SC-20-SC), a
collaborative effort of the U.S. Department of Energy Office of Science and
the National Nuclear Security Administration.

This research used resources of the Oak Ridge Leadership Computing Facility at
the Oak Ridge National Laboratory, which is supported by the Office of Science
of the U.S. Department of Energy under Contract No. DE-AC05-00OR22725.

This research used resources of the Compute and Data Environment for Science
(CADES) at the Oak Ridge National Laboratory, which is supported by the Office
of Science of the U.S. Department of Energy under Contract No.
DE-AC05-00OR22725.
\ifjournal
\end{acks}
\fi

\bibliographystyle{ACM-Reference-Format}
\bibliography{main}

\end{document}